\documentclass[preprintstyle]{ckm} 

\confname{Workshop on the CKM Unitarity Triangle, IPPP Durham, April
  2003}

\title{$K_{l3}$ prospects from NA48/2}

\author{D. Madigozhin for NA48/2 collaboration}
\address{Joint Institute for Nuclear Research, Dubna, Moscow region, 141980}

\begin{document}

\begin{abstract}
NA48/2 plans for the $K_{e3}$ and $K_{\mu 3}$ branching ratios and 
Dalitz plots measurements are discussed. The relative errors for the
branching ratios $< 1\%$ are expected.
\end{abstract}

\maketitle

\section{Motivation}

In the most precise test of the CKM matrix unitarity:

\begin{equation}
 |V_{ud}|^2 + |V_{us}|^2 + |V_{ub}|^2 = 0.9957 \pm 0.002 
\end{equation}

the result is less than 1 by more than an error, and $|V_{us}|$ 
contributes near 30\% of the total uncertainty. So it is important 
to measure  $|V_{us}|$ with the precision better than 1\% to clarify 
the situation. 

The values of $\Gamma(K_{e3})$ and $\Gamma(K_{\mu 3})$ are the
input for the $|V_{us}|$ calculation (they enter as multiplicative factors), 
so the next-generation measurements of these rates would improve the precision 
of the CKM unitarity test. The Dalitz plot shapes are also important, as the
decay form factor contributes here.

\section{NA48/2 experiment}

The main NA48/2 goal is a search for the CP violation in the asymmetry
of  $\pi^+\pi^+\pi^-$ and $\pi^+\pi^0\pi^0$ Dalitz plots between the
kaons of the opposite charge. The additional tasks must not disturb
the main data taking process. 

\begin{figure}
\hbox to\hsize{\hss
\includegraphics[width=\hsize]{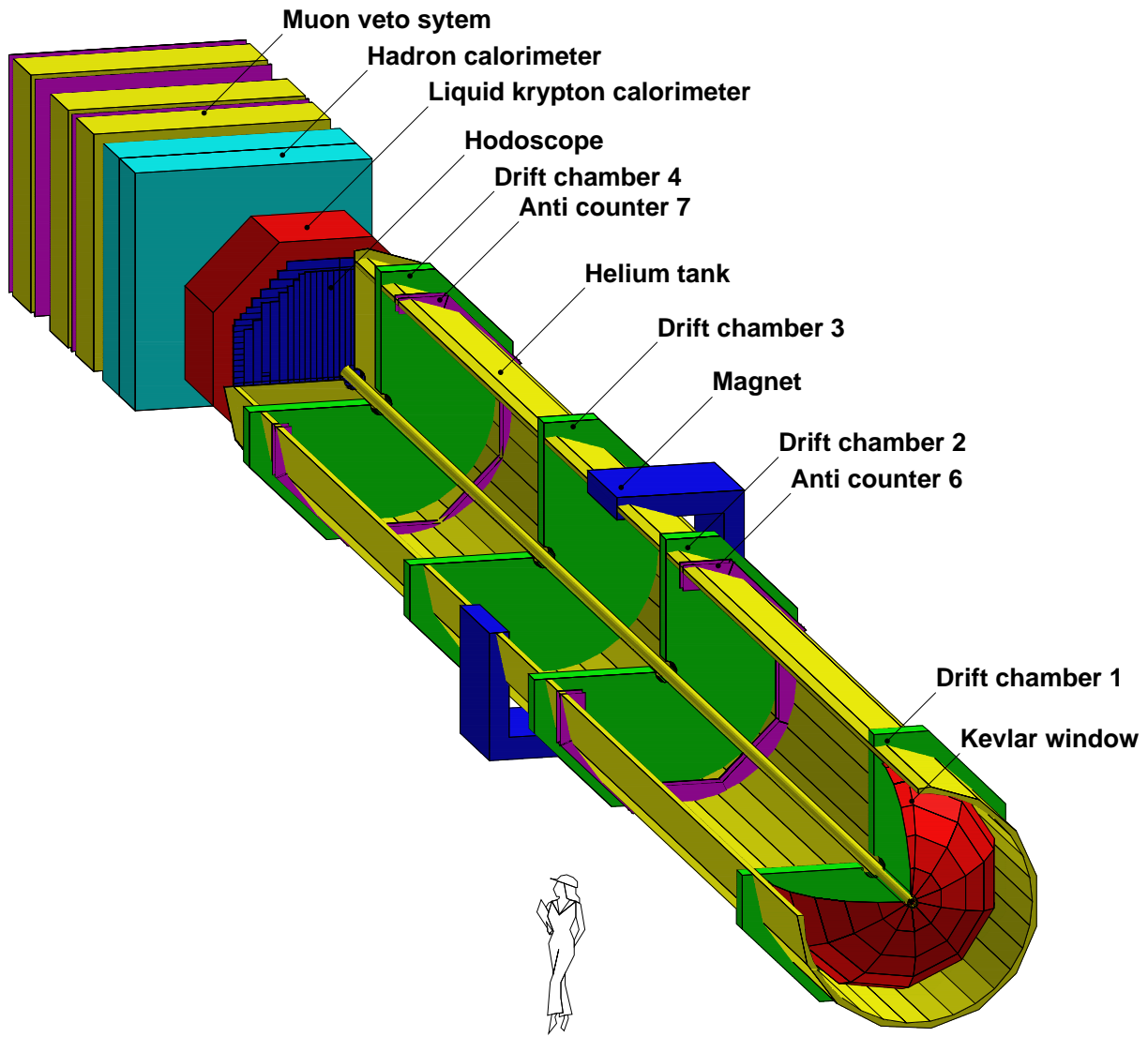}
\hss}
\caption{NA48 detector.}
\label{fig:detector}
\end{figure}

The NA48 detector (fig. \ref{fig:detector}), initially designed for the 
precise measurement of the direct CP violation parameter $Re(\epsilon'/\epsilon)$, 
is described elsewhere \cite{Lai:2001ki}, \cite{Batley:2002gn}. The main 
new features of the NA48/2 stage \cite{Batley:1999fv} setup that may be relevant 
for the $K_{l3}$ measurements are the following:

\begin{itemize}

\item The narrow momentum spectrum ($60 \pm 3$ GeV) of charged kaons. 

\item KAon BEam Spectrometer (KABES), based on the MICROMEGAS technology
\cite{Charpak:2001tp}, placed before the decay volume will 
provide the kaon momentum measurement with $\approx 1\%$ precision. It will
give a possibility to reconstruct the kinematics of the decays with neutrino
among the products.
 
\end{itemize}

\section{Measurement with a similar normalizing modes}

There are several measurement strategies, discussed now in the
NA48/2 collaboration. The choice of the normalizing 
modes will define the expected sensitivity of the measurement
to the results of other experiments (normalizing mode rates) 
and to the quality of the Monte Carlo acceptance calculations.
Moreover, it will define the extra trigger rate that has
to be affordable for NA48/2.

One can normalize the branching measurement with 
the similar decays. In this case we will measure the branching ratios
$\frac{Br(\pi^0 e^+\nu_e)}{Br(\pi^+\pi^0)}$ and 
$\frac{Br(\pi^0 \mu^+\nu_{\mu})}{Br(\pi^+\pi^0)}$.

{\bf Pro:} the same number of $\gamma$-s and charged tracks may decrease
the sensitivity to the Monte Carlo acceptance calculations.     

The error of the measured fraction $Br$ with $B_2$ as normalizing one
is (ignoring the acceptance errors):

\begin{equation}
\delta Br=\frac{N}{N_n} B_2 (\frac{\delta B_2}{B_2} +\sqrt{\frac{1}{N}+\frac{1}{N_n}})
\end{equation}

{\bf Contra:} We can not decrease the relative error of $K_{l3}$ fractions to 
better than 0.007 (PDG error of $Br(\pi^+\pi^0)$) this way. In fact, 
the precision can not be better than 1\% (the statistics doesn't limit here, 
but Monte Carlo uncertainty will contribute).  

One can measure $\frac{Br(\pi^0 \mu^+\nu_{\mu})}{Br(\mu\nu)}$.

{\bf Pro:} Better precision of the $Br(\mu\nu)$ (relative error 0.003). May be 
some cancelling of the Muon Veto efficiency.

{\bf Contra:} $(\mu\nu)$ event doesn't have $\gamma$ and the event is not very 
similar to $K_{\mu3}$ from the detector point of view. We will rely more on the 
precision of the acceptances calculation.
 
\section{Using all large modes for normalizing}

One can record all the large fractions (with proper downscalings),
including ($\mu^+\nu_{\mu}$) {\bf together with the corresponding
radiative processes: $\mu^+\nu_{\mu} \gamma$, $\pi^+\pi^0 \gamma$ etc.}.

One can see that the sum of all the other charged kaon decays (excluding the exotical ones) 
is $0.00011 \pm 0.00013$. This contribution is negligible for our task.

Assuming that the sum of the 6 largest modes is $\approx 1$ , we may calculate the
specific large branching ratio $Br_x$ as follows: 

\begin{equation}
Br_x=\frac{N_x D_x / A_x}{\Sigma N_i D_i / A_i} = (\Sigma \frac{D_i}{D_x} \frac{A_x N_i}{A_i N_x})^{-1}
\label{first}
\end{equation}

Here $N_i$ is the number of reconstructed $i$-th mode decays, $A_i$ is the acceptance of the mode,
$D_i$ is the corresponding downscaling factor. The sum is running over the first 6 modes, 
indexed by $i$. 

The expression for the error is:

\begin{equation}
\frac{\sigma Br_x}{Br_x}=\sqrt{
              (1-Br_x)^2((\frac{\sigma A_x}{A_x})^2+\frac{1}{N_x}) +
\Sigma_{i \neq x} Br_i^2((\frac{\sigma A_i}{A_i})^2+\frac{1}{N_i})
}
\end{equation}

In the very optimistic case the acceptance calculation
precisions may reach per mille level, so to have a compatible statistical error
one needs not more than $10^6$ events for each large mode (or less than 10 
events per burst in each mode, as normally NA48 takes 200 000 - 300 000 bursts
per year).

\section{Possible triggers and statistics}

The preliminary Monte Carlo estimation of the particle fluxes gives the
following results. In the decay volume of 108 meters there will be near 
$1.04 \times 10^6$ charged kaon decays and about 1.75 times more
pion decays. From the test run 2001 year we expect a considerable flux of
muons from target also, but it will be much weaker in 2003 due to the beam 
line development.  
\hspace{1cm}

\begin{tabular}{|l|l|l|l|} \hline
 \# & Mode &                acceptance  & events/burst \\
1 & $\mu^+\nu_{\mu}$      & 76 \% & 503000  \\
2 & $\pi^+\pi^0$          & 22 \% & 48000 \\
3 & $\pi^+\pi^+\pi^-$     & 26 \% & 15000 \\
4 & $\pi^+\pi^0\pi^0$     & 7 \% & 1300 \\
5 & $\pi^0\mu^+\nu_{\mu}$ & 24 \% & 8000 \\
6 & $\pi^0 e^+\nu_e$      & 17 \% & 8500 \\ \hline
\end{tabular}

A large statistics of $\pi^+\pi^+\pi^-$ (billions) and $\pi^+\pi^0\pi^0$ will 
provide a chance to make a precise tuning of the Monte Carlo (beam geometry, 
acceptances, spectra).

Maximum trigger readout rate is estimated to be 60K in 5.2 s spill. Main triggers
for $3\pi$ decays may use 50-55 K (the purity of the trigger for $\pi^+\pi^+\pi^-$ 
in the worst case is 0.4, for $\pi^+\pi^0\pi^0$ the expected trigger rate is 16K).  
So only few thousand triggers could be used for other purposes.

If we consider a rather conservative scenario of data taking (simple, but efficient 
triggers with proper downscalings), we can collect more than 20 good events for 
each of the above modes, taking less than 1700 extra triggered events in total. 
With more advanced trigger logics (almost ready at present) we will collect 10 
times more $K_{l3}$ events to mesure the Dalitz plot shapes, taking less than 2000 extra 
triggers.

\section{Conclusion}

\begin{itemize}

\item The $K_{e3}$ and $K_{\mu 3}$ branching ratios and Dalitz plots
measurement is foreseen in the charged kaon beam of NA48/2 experiment.

\item The dominating contribution to the uncertainties may be 
systematic rather than statistical error. Assuming the reasonable
level of the Monte Carlo error one can set a limit on the statistics
we really need. It is $10^6$ per normalizing mode and $10^7$ per 
$K_{l3}$ mode (10 times more for Dalitz plots). 

\item The data can be taken without disturbing the main NA48/2 tasks. 
The additional output triggers rate will be less than 3\% of the total 
one.  

\item The relative branching error is expected to be between 0.01 and 0.001, 
depending on the Monte Carlo precision.

\end{itemize}

\end{document}